\documentstyle[epsf,aps,prl]{revtex}
\newcommand{\p}[2]{\frac{\partial #1}{\partial #2}}
\newcommand{\pp}[2]{\frac{\partial^2 #1}{\partial #2^2}}

\newcommand{\od}[2]{\frac{d #1}{d #2}}
\newcommand{\pdiv}[1]{\nabla\cdot #1}
\newcommand{\grad}[1]{\nabla #1}

\newcommand{\al}{\alpha}

\newcommand{\ve}{{\bf E}}
\newcommand{\vet}{\tilde{\bf E}}
\newcommand{\vb}{{\bf B}}
\newcommand{\vbt}{\tilde{\bf B}}
\newcommand{\vv}{{\bf V}}
\newcommand{\vx}{{\bf r}}
\newcommand{\vk}{{\bf k}}

\newcommand{\be}{\begin{equation}}
\newcommand{\ee}{\end{equation}}
\newcommand{\ba}{\begin{eqnarray}}
\newcommand{\ea}{\end{eqnarray}}
\begin{document}

\twocolumn[\hsize\textwidth\columnwidth\hsize\csname@twocolumnfalse\endcsname 
\title{Dissipative effect of thermal radiation loss in high-temperature dense
plasmas}
\author{L H Li\dag\ddag\ and H Q Zhang\dag}
\address{\dag\ Purple Mountain Observatory, Academia Sinica,
 Nanjing, Jiangsu 210008, China}
\address{\ddag\ The Institute of Physical and Chemical Research, 2-1, 
 Hirosawa, Wako, Saitama, 351-01, Japan}
\maketitle
\begin{abstract}
A dynamical model based on the two-fluid dynamical equations with energy
generation and loss is obtained and used to investigate the
self-generated magnetic fields in high-temperature dense plasmas such
as the solar core. The self-generation of magnetic fields might be
looked at as a self-organization-type behavior of stochastic thermal
radiation fields, as expected for an open dissipative system according
to Prigogine's theory of dissipative structures.
\end{abstract}
\pacs{PACS 52.25.Nr, 52.35.Ra}
] \renewcommand{\thefootnote}{\arabic{footnote}} \setcounter{footnote}{0}

Thermal motion of electrons generates thermal radiation
mainly by means of thermal bremsstrahlung emission in 
high-temperature plasmas. The higher the plasma temperature the more intense 
thermal radiation of the plasma. Thermal radiation causes substantial
energy loss in high-temperature plasmas such as thermal nuclear fusion
plasmas; for instance, the well-known Lawson condition is derived from
the condition that the fusion energy compemsates the thermal bremsstrahlung 
emission loss. Here we are about to report some dissipative effect of
thermal radiation loss in high-temperature dense plasmas such as thermal
nuclear fusion plasmas.

Generally, the elastic collision frequency $\nu_{ei}$ between electrons
and ion is high in high-temperature dense plasmas, for example, in the
core of the Sun $T_e\sim 10^{7}$ K, $n_e\sim 10^{32}\ m^{-3}$, we can
eastimate $\nu_{ei}\sim 10^{16}$ Hz, the plasma frequency $\nu_{pe}\sim
10^{17}$ Hz. Such a frequent collision would destroy any collective plasma
motion or coherent structure if the Sun were isolated adiabatically. In
other word, the Sun would be in thermodynamic equilibrium if it was
an isolated system. However, the Sun is in fact an open system: it not
only gains thermal energy from the inelastic collisions between ions
such as the proton-proton chain (or pp chain), but also loses energy by
means of thermal radiation, for example, the loss rate of thermal radiation or
the photon luminosity $L_{\odot} \sim 4\cdot 10^{33}$ erg/s (Bahcall, 1989).
Therefore, the Sun may be far from equilibrium. The theory of dissipative
structures (Prigogine, 1973; 1977) has shown that coherent behaviors may occur
in a dissipative system. In fact, we have witnessed and are witnessing the
dissipative structures in the convectve zone of the Sun, the convective cells.
These are the coherent structures at the outer part of the Sun caused by
thermal radiation dissipation which maintains the necessary temperature
gradient. What are the dissipative structures in the source region?

Roughly, nuclear fusion reactions take place within $0.3\/ R_{\odot}$
(Bahcall, 1989), where nuclear reactions (inelastic collisions between
ions) first enhance stochastic motion of ions, the ions then transfer
energy and momentum to electrons via elastic collision. The electrons
lose energy and momentum through emitting photons via inelastic
collision with ions and the thermal radiation escapes from the source
region via radiation transfer. On average, the direction of the energy
and momentum flux in the nuclear fusion region is therefore:
\be
  \mbox{ions}\rightarrow \mbox{electrons}
    \rightarrow \mbox{photons} \label{nequi}
\ee
This implies that electrons do not reach thermodynamic equilibrium
with ions ($T_i>T_e$), nor do photons with electrons in the fusion
plasma and that it is the very ions that replenish the radiation loss of
both energy and momentum of electrons.

This nonequilibrium effect also shows itself in the two-fluid dynamical
equations (Rose and Clark, 1961; Ma {\em et al}, 1988; Rybicki and
Lightman, 1979):
\be
  \p{n_{\al}}{t}+\pdiv{(n_\al\vv_\al)}=\frac{\delta
     n_\alpha}{\delta t}   \label{macro1}
\ee
\be
  m_\al n_\al\od{\vv_{\al}}{t}=-\pdiv{{\bf P}_\al}+n_\al{\bf F}_\al
  +{\bf R}_\al+\delta{\bf R}_\al
   \label{macro2}
\ee
\be
\frac{3}{2}n_\al\od{T_\al}{t}=-({\bf P}_\al\cdot\grad)\cdot\vv_\al
  -\pdiv{{\bf q}_\al}+Q_\al+\delta Q_\al 
  \label{macro3}
\ee
\be
  \od{I_\nu}{s}=-\al_\nu I_\nu+j_\nu  \label{transfer}
\ee
where $ds$ is a differential element of length along the ray, $I_\nu$ is
the energy crossing unit area in unit time, unit solid angle and unit
frequency range of radiation fields, called as the specific intensity or
brightness, $\al_\nu$ and $j_\nu$ are the absorption coefficient and
emission coefficient respectively. ${\bf P}_\al$ is called the kinetic
stress tensor, ${\bf q}_\al$ the thermal flux vector, ${\bf F}_\al$ is
the total force exerted on the fluid element of $\al$ species,  Of five
collision-related terms, ${\delta n_\al}/{\delta t}$ represents
created or destroyed particle density of $\al$ species in the unit time,
${\bf R}_\al$ is the momentum gain or loss rate of the fluid element of
$\al$ particles by their elastic collision with the unlike particles
(including photons), the same for ${\delta\bf R}_\al$ but due to the
inelastic collision. Similarly, $Q_\al$ and $\delta Q_\al$ are thermal
energy gain or loss rate due to the elastic collision and the inelastic
collision respectively.
 
For the sake of clarity, we assume that the quasi-stationary state has
been reached with $\partial{T_\alpha}/\partial{t}\approx 0 $, as
believed in the solar core. If we further neglect viscosity, thermal
conduction, thermal convection and restrict to a very short spacescale
($<1 cm$), the energy conservation equation reduces to
\be
 0=Q_\al+\delta Q_\al   \label{energy}
\ee
Eq.(\ref{transfer}) tells us that radiation energy is continuously
transfered from the nuclear fusion region to the outside due to the
radiation loss, which implies that electrons lose energy through
emitting photon via inelastic collision with ions, $\delta Q_e<0$,
Eq.(\ref{energy}) thus leads to $Q_e=-\delta Q_e>0$. This means that
electrons should gain energy on average from ions via elastic collision
with ions. In the quasi-stationary case, $Q_i\equiv\sum_{\al\ne
e}Q_\al=-Q_e<0$, which implies that ions lose energy on average via
elsatic collisions with electrons. The energy loss of ions $Q_i<0$ will
be replenished by the energy generation $\delta Q_i>0$ due to the
inelastic collision (i.e., nuclear reaction) between ions or ions and
electrons, $Q_i=-\delta Q_i$, as implied by the energy conservation
equations for ions. The energy transfer accompanies similar momentum
transfer between electrons and ions: ${\bf R}_e=-{\bf
R}_i\equiv\sum_{\al\ne e}{\bf R}_\al>0$, which means that ions lose
momentum and electrons gain momentum on average in the elastic collision
between ions and electrons in the quasi-stationary case. The momentum
generation $\delta {\bf R}_i\equiv\sum_{\al\ne e}\delta{\bf R}_\al>0$ of
ions due to the nuclear fusion collision replenishes the momentum loss
${\bf R}_i<0$ of ions. 

The following task is to estimate ${\bf R}_i+\delta{\bf R}_i$ and ${\bf
R}_e+\delta{\bf R}_e$. In so doing we have to take into account the
radiation fields. The transfer equation (Rybicki and Lightman, 1979)
takes a particularly simple form if, instead of $s$, we use the optical
depth $\tau_\nu$ defined by $d\tau_\nu=\al_\nu ds$, 
\be
  \od{I\nu}{\tau_\nu}=-I_\nu+j_\nu/\al_\nu
\ee
In plane-parallel media, a standard optical depth is sometimes used to
measure distance normal to the surface, so that $\tau_\nu=\tau_\nu(z)$.
A medium is said to be optically thick or opaque when $\tau_\nu$,
integrated along a typical path through the medium, satisfies
$\tau_\nu>1$. An optically thick medium is one in which the average
photon of frequency $\nu$ cannot traverse the entire medium without
being absorbed. Solar core plasma is optically thick, and the absorption
is mainly due to the (massive) ions (Bacall, 1989). If a medium absorbs
radiation, then the radiation exerts a force on the medium, because
radiation carries momentum. The radiation pressure ${\bf F}_{rad}$ can
be defined by the specific intensity $I_\nu$ in remembering that a
photon has momentum $h\nu/c$,
\be
  {\bf F}_{rad}=\frac{1}{c}\int\al_\nu I_\nu {\bf n}d\Omega d\nu
\ee
where ${\bf n}$ is a unit vector along the direction of the ray,
$z$-direction, for instance. Since the absorption is mainly due to the
ions, we know
\be
  {\bf R}_i+\delta{\bf R}_i={\bf F}_{rad}  \label{ion_m}
\ee
\be
  {\bf R}_e+\delta{\bf R}_e\approx 0    \label{electron_m}
\ee
As a result, the momentum transport equation reduces to
\be
  m_i n_i\od{\vv_i}{t}=-\grad{p_i}+n_i{\bf F}_i+{\bf F}_{rad}
     \label{momentum_i} 
\ee
\be
  m_e n_e\od{\vv_e}{t}=-\grad{p_e}+n_e{\bf F}_e
     \label{momentum_e} 
\ee
where $m_i$, $n_i$ and $\vv_i$ are the effective mass, density of ions
and the effective velocity of an ion-fluid element. This is in agreement
with the nonequilibrium processes (\ref{nequi}). The most outstanding
characteristic of the reduced equation is that the collision terms have
been canceled out. In contrast, the cancellation cannot occur in the
equilibrium case because there is no a well-determined direction of
energy and momentum flux. Consequently, we find out that the
nonequilibrium effect cancels the collision effect.

In the two-fluid approximation $\vv_\al=\vv_\al(\vx,t)$ is the velocity
of an $\al$ fluid element, which is the first-order
moment of the distribution function at the velocity space of the $\al$
particle. Under such an approximation, the acceleration behavior of
single particles has been ruled out. Therefore, the
Maxwell equations (Rose and Clark, 1961; Ma {\em et al}, 1988)
\be
  M(\ve,\vb;\rho,{\bf J})=0 \label{maxwell}
\ee
with induced charge density $\rho=\sum_{\al}q_\al n_\al$ and induced current
density ${\bf J}=\sum_\al n_\al q_\al\vv_\al$ determine only the induced fields
in the plasma by motion of fluid elements. The radiation fields
$\ve_{rad}$ and $\vb_{rad}$ are generated by the acceleration motion of
single particles and thus can be determined only by the retarded
potentials of single moving charges, or Li\'enard-Wiechart potentials.
On averaging single particle fields $\ve_{sin}$ and $\vb_{sin}$ in the
velocity space with the distribution function of electrons as the weight
factor, we can obtain the radiation fields (Rybicki and Lightman, 1979).
At this stage, we are able to write down explicitly
\be
 {\bf F}_\al={q_\al}[(\ve+\ve_{rad})+\vv_\al\times(\vb
  +\vb_{rad})]+{\bf g}_\al
\ee
where ${\bf g}_\al$ is gravity, $\al=i,\ e$. The quasi-stationary condition
requires that the thermal and radiation pressure balance the gravity on the
macroscopic level ($\gg 1cm$),
\be
  -\nabla p_i+{\bf F}_{rad}+{\bf g}_i=0  \label{balance_i}
\ee
\be
  -\nabla p_e+{\bf g}_e=0  \label{balance_e}
\ee
As a result, Eqs.(\ref{momentum_i}) and (\ref{momentum_e}) reduce to
\be
  m_\al n_\al\od{\vv_{\al}}{t}=-\gamma_\al T_\al\grad{\tilde{n}_\al}+{n_\al
    q_\al}(\ve^{tot}+\vv_\al\times\vb^{tot}) 
    \label{momentum1}
\ee
where $\tilde{n}_\al$ ($\al=i,\ e$) is the microscopic density
fluctuation of the $\al$ fluid. $\ve^{tot}=\ve+\ve_{rad}$, the same for 
$\vb^{tot}$. We have assumed thermal pressure $p_{\al}=\gamma_\al n_\al
T_\al$ with the polytropic index $\gamma_\al$. The mass consernvation
equation can be rewritten down as follows by neglecting particle
creation or destruction:
\be
  \p{n_{\al}}{t}+\pdiv{(n_\al\vv_\al)}=0 \label{mass}
\ee

So far we have shown that the plasma in the nuclear fusion region can
approximately be described by Eqs.(\ref{maxwell}), (\ref{momentum1}) and 
(\ref{mass}) when the energy and momentum flux (\ref{nequi}) is
well-defined by the thermal radiation loss. These equations possess two
characteristics: (a) The elastic collision terms have been canceled out
by the inelastic collision terms; (b) thermal radiation fields only
appear in the momentum transport equations. The first confirms the fact
that non-equilibrium may be a source of order, the second allows us to
study the self-organization phenomena (Nicolis and Prigogine, 1977) of
the random thermal radiation fields. Self-organization will lead to
order, or dissipative structures.

Self-organization needs nonlinearity. The reduced two-fluid dynamical
equations contain the needed nonlinearity. These equations can be
simplified substantially through distinguishing the ion-timescale
$\tau_i\sim \omega_{pi}^{-1}$ and the electron-timescale
$\tau_e\sim\omega_{pe}^{-1}$ with the self-generated effect of magnetic
fields included in (Kono {\em et al}, 1981; Li, 1993a,b):
\be
  i\p{\ve}{t}+\alpha\pp{\ve}{z}-(\beta z+pn)\ve+ip\ve\times\vb
    ={\bf S}
\ee
\be
  i\p{E_z}{t}+\pp{E_z}{z}-(\beta z+pn)E_z+ip[\ve\times\vb]_z
    =S_z
\ee
\be
  (\pp{}{t}+2\nu_i*\p{}{t}-\gamma\pp{}{z})n
    =\gamma'\pp{}{z}{|\ve|^2}
\ee
where $\ve$ is the slowly varying complex amplitude of the
induced high-frequency ($\omega_0$) electric field $\tilde{\ve}$:
$\tilde{\ve}=\frac{1}{2}\{\ve\exp(-\omega_0 t)+c.c.\}$. The plasma has
been assumed to be plane-parallel to the density gradient of
scalelength L along z-axis. Therefore, all variables depend on only z.
Then,  $\vk=(0,0,k)$. $\ve^t=[E_x, E_y]$ is the transverse field
component, while $E_z$ is the longitudinal field component. The validity
conditions of these equations are $k/k_{De}\ll 1$, $W k/k_{De}\ll 1$ and
$|\delta n|/n_e\le\delta\nu/\nu_{pe}\ll 1$. Another physical constraint
is  $\delta n/n_e\le \delta\nu/\nu_{pe}$. All variables have been
rescaled as follows (Morale and Lee, 1977): 
$\vb=\tilde{\vb}_s/B_{ML}$, $\ve=\tilde{\ve}/E_{ML}$,
$n=(\delta{n}/n_e)/N_{ML}$, $t=\tilde{t}/T_{ML}$ and
$z=\tilde{z}/Z_{ML}$, in which the tilded stand for the unrescaled.
$B_{ML}=\beta_1a^2W$, $E_{ML}=2\beta_1a\tilde{E}_0$,
  $N_{ML}=\beta_1a^2W$, 
  $T_{ML}=2a/\beta_1\omega_{pe}$ and
  $Z_{ML}=a^{-1}L/\sqrt{\beta_1}$ are the new units, in which
  $a=(L\omega_{pe}/\sqrt{3}V_{Te})^{2/3}$,
  $\beta_1=1+3(T_i/T_e)$, $W=\tilde{E}_0^2/(n_e\kappa T_e/\epsilon_0)$.
The other parameters are: $\alpha=(c/V_{Te})^2/3$,
$\beta=\beta_1^3$, $p=a^3W/(1-B_0^2)$ in that
$B_0=\tilde{B}_s/(m_e\omega_{pe}/e)$, $\gamma=\frac{4}{3}a^3(m_e/m_i)$,
$\gamma'=\gamma/(1-B_0^2)^2$, $S_x=(1/2\beta_1^2)\cos\phi\sin\theta$,
$S_y=(1/2\beta_1^2)\sin\phi\sin\theta$ and $S_z=(1/2\beta_1^2)\cos\theta$
in that both $\theta$ $\in$ $[0,\pi]$ and $\phi$ $\in$ $[0,2\pi]$ are random
due to random orientation of the thermal radiation field.
$\tilde{E}_0\approx[\rho(\nu_{pe},T)\delta\nu/\epsilon_0]^{1/2}$  
is the thermal radiation field components at and near the local plasma
frequency $\nu_{pe}$, where $\rho(\nu,T)$ is spectral energy density
of thermal radiation. The symbol ``*'' means convolution product,
$\nu_i$ is the Landau damping of low-frequency longitudinal waves.

The so-called self-generated magnetic field $\vbt_s$ of the plasma is
the slowly varying component of the induced magnetic field (Kono {\em et
al}, 1981):
\be
  \frac{\vbt_s}{B_c}=(\frac{V_{Te}}{c})^2\frac{i\vet\times\vet^*}{E_c^2}
     \label{magnet}
\ee
where $B_c=m_e\omega_{pe}/e$ is the critical magnetic field at which electron 
gyrofrequency $\Omega_e=eB_c/m_e$ equals to local plasma (angular) frequency 
$\omega_{pe}$, $E_c=(n_e\kappa T/\epsilon_0)^{1/2}$. The magnetic-field
generation was shown to be due to a solenoidal current ${\bf
j}_s=-ie\omega_{pe}^2/(16\pi m_e\omega_0^3)\nabla\times(\ve\times\ve^*)$
[Eq.(1.1) of Kono {\em et al}, 1981] by many authors (see Kono {\em et al},
1981 and references cited therein). Through taking the curl of the
solw-timescale Amp\'ere law and neglecting the displacement current, one
obtains [Eq.(4.7) of Kono {\em et al}, 1981]
\be
  \nabla\times\nabla\times\vbt_s=\frac{4\pi}{c}\nabla\times{\bf j}_s
\ee
This equation reduces to Eq.(\ref{magnet}) when $\nabla\cdot\vbt_s=0$ is
satisfied, as one of the Maxwell equations requires. Eq.(\ref{magnet})
shows that if there is no self-organization of the stochastic thermal
radiation field $\ve_{rad}$, which has been incorporated in parameters
${\bf S}$ and $p$, the self-generated magnetic field $\vbt_s$ must be
very weak, vice versa. Therefore, we can investigate the
self-organization effect of the thermal radiation field through
monitoring $\vbt_s$. In order to make sure that we have observed the
self-organization phenomenon, we have to use the statistical average
$\tilde{B}_0(\tilde{z})$ of $\tilde{B}_s(z',\tilde{t})$ over time, space
and random initial conditions: 
\be
  \tilde{B}_0(\tilde{z})=\frac{1}{n-2}\left(\sum_{i=1}^n 
    \overline{B}_i-\overline{B}_{max}
    -\overline{B}_{min}\right)
\ee
where $\tilde{B}_i=L_{sim}^{-1}\{\int_z^{z+L_{sim}}[\tilde{B}_x^2(z')
+\tilde{B}_y^2(z')]dz'\}^{1/2}$ and $n$ is the number of solutions starting
from random initial conditions. $\overline{B}$ means averaging
$\tilde{\bf B}$ in the range $\tilde{B}_{max}/B_c \in [0.8,0.9]$, where
$\tilde{B}_{max}=Max\{[\tilde{B}_x^2(z)+\tilde{B}_y^2(z)]^{1/2}\}\le
0.9 B_c$ is one of our controlling parameters (the other is the maximum
density fluctuation $\delta n/n_e\le \delta\nu/\nu$). $L_{sim}$ is the
simulation cell size.

\begin{figure}
\hskip-1.1truecm{\epsfxsize=3.6 in \epsfbox{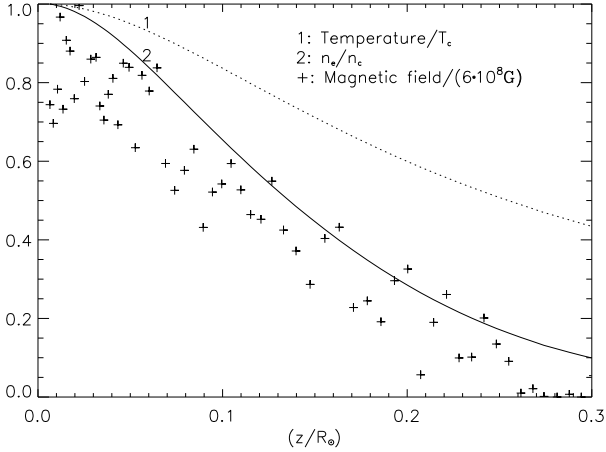}}
\caption{Variation of the self-generated magnetic field with the solar
radius coordinate $z/R_\odot$, where $R_\odot$ is solar radius.
Curves~1 and 2, representing solar plasma temperature and electron
number density respectively, are the input parameters. Symbols ``+''
stand for the corresponding self-generated magnetic fields.}
\label{fig:magnet}
\end{figure}

A standard second-order explict quasispectral method (Li and Li, 1993;
Li, 1996) has been used to numerically solve the model equations in
order to calculate $B_x$ and $B_y$ (note: $B_z=0$) and thus
$\tilde{B}_s$. We have used table XVI of Bahcall \&\
Pinsonneault's standard solar model (1992) to calculate our input
parameters: the electron and ion temperatures $T_e\approx T_i=T$, electron
number density $n_e=\frac{1}{2}(1+X)\rho/m_H$, and plasma scalelength
$L=n_e(dn_e/d z)^{-1}$, where $X=X(^1H)+X(^3He)$
represents the concentration, by mass, of hydrogen. Although the
simulation cells are small, $L_{sim}\sim 10^{-3}$ m, it is much larger
than the Debye length $\lambda_{De}=V_{Te}/\omega_{pe}\sim 10^{-11}$ m.
Figure~\ref{fig:magnet} shows $\tilde{B}_0(z)$ obtained by using $n=10$ (the
convergence trend has been found by using $n=1,4,8$ and 10), which shows
at least $0.4$ gigagauss magnetic field may be generated at the center
of the Sun. Such a strong field shows that the self-organization
behavior of the stochastic thermal radiation fields does occur. 
The collision time scale is about $10^{-16}$ s, while the growth time
scale of the self-generated magnetic field is about $10^{-12}$ s. The
fact that the coherent time scale is much longer than the collision time
scale is also an indicator of self-organization (Nicolis and Prigogine,
1977).

\vskip0.3truecm

The authors gratefully acknowledge the anomymous referee for his/her
helpful comment and one of the authors (Li) wants to thank M Matsuoka
for the kind hospitality met during my year-long stay at RIKEN, where
this work was partly completed.

\section*{References}

\newdimen\bindent
\bindent=2em
\newcommand{\bib}{\par\hangindent=\bindent}
\newcommand{\BHarvard}{\begingroup\parindent=0pt}
\newcommand{\EHarvard}{\par\endgroup}

\BHarvard

\bib Bahcall J N 1989 {\em Neutrino Astrophysics} (Cambridge:
  Cambridge University Press)

\bib Bahcall J N and Pinsonneault M H 1992 {\em Rev. Mod.
  Phys.\/}~{\bf 64} 885 

\bib Kono M, \v{S}koric M M \&\ ter Haar D 1981 {\em J. Plasma
  Phys.\/}~{\bf 26} 123

\bib Li L H 1993a {\em Phys. Fluids B\/}~{\bf 5} 350

\bib Li L H 1993b {\em Phys. Fluids B\/}~{\bf 5} 1760

\bib Li L H 1996 {\em J. Phys. D:\/ Appl. Phys.\/}~{\bf 29} 267

\bib Li L H and Li X Q 1993 {\em Phys. Fluids B\/}~{\bf 5} 3819

\bib Ma T C, Hu X W and Chen Y H 1988 {\em Principles of Plasma Physics}
  (Hefei: U C S T Press)

\bib Morales G J and Lee Y C 1977 {\em Phys. fluids\/}~{\bf 20} 1135

\bib Nicolis G and Prigogine I 1977 {\em Self Organization in
  Nonequilibrium Systems} (London: Wiley J)

\bib Prigogine I 1973 in {\em The Physicist's Conception of Nature} 
  (Dordrecht: Reidel D) 

\bib Rose D J and Clark M Jr 1961 {\em Plasmas and Controlled Fusion},
  (New York: The M I T Press \&\ Wiley J)

\bib Rybicki G B and Lightman A P 1979 {\em Radiative Processes in
  Astrophysics} (New York: Wiley J \&\ Son)

\EHarvard

\end{document}